\documentstyle[twoside]{article}
\newcount\Mac  \Mac=0  % devo mettere Mac=1 se sto lavorando sul file Mac
\newcommand{\ifMac}[2]{\ifnum\Mac=1 #1 \else #2 \fi}
\newcommand{\riga}[1]{\noalign{\hbox{\parbox{\textwidth}{#1}}}\nonumber}
\newcommand{\bAk}[3]{\langle #1|#2|#3\rangle}
\newcommand{\One}{\hbox{1\kern-.24em I}}

\newcommand{\Opx}[1]{{\cal O}^{\times}_{#1}}
\newcommand{\Opi}[1]{{\cal O}^{=}_{#1}}

\newcommand{\NP}{Nucl. Phys.}
\newcommand{\PRL}{Phys. Rev. Lett.}
\newcommand{\PL}{Phys. Lett.}
\newcommand{\PR}{Phys. Rev.}

\newcommand{\mbR}{m_{\tilde{b}_R}} \newcommand{\mbL}{m_{\tilde{b}_L}}

\newcommand{\rbL}{\ell_3}          \newcommand{\rbR}{r_3}
\newcommand{\rdL}{\ell_{12}}    \newcommand{\rdR}{r_{12}}
\newcommand{\BM}{B_{-\!\!-\hspace{-1.8ex}\times\,\,}}
\newcommand{\Bk}{B_{-\!\!-\hspace{-1.8ex}>\,\,}}
\newcommand{\Pglu}{P_{-\!\!-\hspace{-1.8ex}>\,\,}}
\newcommand{\PgluE}{P_{-\!\!-\hspace{-1.8ex}\times\,\,}}
\newcommand{\PgluI}{P_{\cdots\hspace{-2ex}\cdots\hspace{-1.3ex}\times\,\,}}
\newcommand{\eq}[1]{~{\rm (\ref{eq:#1})}}
\newcommand{\sys}[1]{~{\rm (\ref{sys:#1})}}

\newcommand{\MGUT}{M_{\rm G}}
\def\Red{}
\def\Black{}
\def\Blue{}

\newcommand{\lascia}[1]{}
\makeatletter
\def\art{\@ifnextchar[{\eart}{\oart}}
\def\eart[#1]#2#3#4#5#6{{\rm #2}, {\em #3 \bf #4} {\rm (#6) #5}}
\def\hepart[#1]#2{{\rm #2, \em#1}}
\newcommand{\oart}[5]{{\rm #1}, {\em #2 \bf #3} {\rm (#5) #4}}
\newcommand{\y}{{\rm and} }

\newcounter{alphaequation}[equation]
\def\thealphaequation{\theequation\hbox to
0.6em{\hfil\alph{alphaequation}\hfil}}
\def\eqnsystem#1{
\def\@eqnnum{{\rm (\thealphaequation)}}
\def\@@eqncr{\let\@tempa\relax \ifcase\@eqcnt \def\@tempa{& & &} \or
  \def\@tempa{& &}\or \def\@tempa{&}\fi\@tempa
  \if@eqnsw\@eqnnum\refstepcounter{alphaequation}\fi
\global\@eqnswtrue\global\@eqcnt=0\cr}
\refstepcounter{equation} \let\@currentlabel\theequation \def\@tempb{#1}
\ifx\@tempb\empty\else\label{#1}\fi
\refstepcounter{alphaequation}
\let\@currentlabel\thealphaequation
\global\@eqnswtrue\global\@eqcnt=0 \tabskip\@centering\let\\=\@eqncr
$$\halign to \displaywidth\bgroup \@eqnsel\hskip\@centering
$\displaystyle\tabskip\z@{##}$&\global\@eqcnt\@ne
\hskip2\arraycolsep\hfil${##}$\hfil& \global\@eqcnt\tw@\hskip2\arraycolsep
$\displaystyle\tabskip\z@{##}$\hfil
\tabskip\@centering&\llap{##}\tabskip\z@\cr}

\def\endeqnsystem{\@@eqncr\egroup$$\global\@ignoretrue} \makeatother

\def\Ord{{\cal O}}

\def\circa#1{\,\raise.3ex\hbox{$#1$\kern-.75em\lower1ex\hbox{$\sim$}}\,}

\begin{document}
\begin{quote}
{\em 25 April 1997}\hfill {\bf IFUP--TH 16/97}\\
\phantom{.} \hfill{\bf hep-ph/9704404}
\end{quote}
\bigskip
%\vspace{1cm}
\centerline{\huge\bf\Red CP violation in $B\hspace{-1.7ex}B$ decays}
\centerline{\huge\bf\Red and supersymmetry}
\bigskip\bigskip\Black%\vspace{1cm}
\centerline{\large\bf Riccardo Barbieri {\rm and} Alessandro Strumia} \vspace{0.3cm}
\centerline{\em Dipartimento di Fisica, Universit\`a di Pisa {\rm and}}
\centerline{\em INFN, sezione di Pisa,  I-56126 Pisa, Italia}\vspace{0.3cm}
\bigskip\bigskip\Blue
%\vspace{1cm}
\centerline{\large\bf Abstract}
\begin{quote}\large\indent
CP violation in hadronic $B$-decays is studied in a definite and well motivated
framework for flavour physics and supersymmetry. Possible deviations from the standard
model both in mixing and in decay amplitudes are discussed. An attempt is made to
describe an experimental strategy for looking at these deviations and for measuring the
relevant parameters.
\end{quote}\Black

\section{Introduction}
Supersymmetry may lead to deviations from the expectations
of the Standard Model in a few crucial observables in flavour physics.
Other than the usual Cabibbo-Kobayashi-Maskawa matrix in charged current weak
interactions, in a generic supersymmetric model there are in fact at least two new
possible sources of significant flavour violation.
One is the flavour violating interaction driven by the top Yukawa coupling to the
charged Higgs, with its supersymmetric counterpart.
The other is a possible flavour violation,
originated by misaligned fermion and sfermion mass matrices,
that appears in gaugino (and higgsino)
interactions of matter supermultiplets.

The nature of these two new sources of flavour violation is very different.
The first is there in any realistic supersymmetric extension of the SM,
it is essentially unrelated to supersymmetry breaking and its flavour structure
is controlled by the same CKM matrix.
Its effects may be significant in some loop-induced $B$ decays~\cite{1},
like $b\to s\gamma$ or $b\to s \ell\bar{\ell}$,
but are less important in $\Delta B=2$ or $\Delta S=2$ transitions.
In any event it will not introduce any new CP-violating phase nor it will affect leptons.

The second potential source of flavour violation, on the contrary, arises from new
CKM-like mixing matrices $W$ occurring
in the gaugino ($\lambda$) interactions
\begin{equation}\label{eq:1}
\tilde{\rm f}^\dagger_{L}  W^{\phantom{*}}_{\tilde{\rm f}_{L}
{\rm f}_{L}} {\rm f}_{L} \lambda+
\tilde{\rm f}^\dagger_{R}  W^*_{\tilde{\rm f}_{R} {\rm f}_{R}}
{\rm f}_{R} \lambda+
\hbox{h.c.}
\end{equation}
of matter fermions of given charge, ${\rm f} = {\rm u,d}$ or e, and
chirality, $L$ or $R$, with their superpartners $\tilde{{\rm f}}$.
The presence of a non trivial $W$-matrix in\eq{1} has everything to do with
supersymmetry breaking, since, in the supersymmetric limit, $W=\One$. In particular, it
is essential that the supersymmetry breaking masses of the scalars
$\tilde{{\rm f}}$ be non degenerate in generation space, since, otherwise, any
non-trivial mixing matrix can be rotated away.

There are at least two independent physical motivations that
lead to a particular realization
of\eq{1} which is of interest in this
paper\footnote{For alternatives, see ref.~\cite{NR} and
references therein.}.
One case~\cite{2,3} is supersymmetric unification with a hardness
scale, defined as the highest
scale at which supersymmetry breaking masses appear as local interactions,
%, $\Lambda_H$,
higher than the unification scale $\MGUT$.
The other is the presence of a flavour symmetry, e.g.\ a U(2) symmetry~\cite{4,5},
that might relate the flavour structure
of the fermion mass matrices to the one of the scalar mass matrices.
As pointed out in previous work~\cite{3,5,6},
from this physical picture we expect possibly
significant deviations from the SM in lepton flavour violating processes
(mostly $\mu\to e\gamma$ and $\mu\to e$ conversion in atoms),
in Electric Dipole Moments (EDMs) of the electron and of the neutron,
in CP-violation in the $K$-system and, finally, in
mixing and CP-violation in the $B$-system.
This last issue is the subject of this paper.

\section{The framework defined}
For the present purposes, the physical situation can be characterized as follows:
\begin{itemize}

\item [i)] The d-type quarks of given chirality,
${\rm d} = \{d,s,b\}$, and their superpartners $\tilde{\rm d}$ have gluino
and photino interactions of the form\eq{1} with $W$ matrix elements
of the order of the corresponding CKM matrix elements.
Of special interest are four parameters, independent from
phase conventions in the physical d-quark basis,
\begin{equation}\begin{array}{ll}\displaystyle
\omega_{d_L} \equiv \frac{W_{\tilde{b}_Ld_L}^* W_{\tilde{b}_L b_L}^{\phantom{*}}}
{V_{td}^*V_{tb}^{\phantom{*}}},\qquad &
\displaystyle
\omega_{d_R} \equiv \frac{W_{\tilde{b}_Rd_R}^* W_{\tilde{b}_R b_R}^{\phantom{*}}}
{V_{td}^*V_{tb}^{\phantom{*}}}\\[5mm]
\displaystyle
\omega_{s_L} \equiv \frac{W_{\tilde{b}_Ls_L}^* W_{\tilde{b}_L b_L}^{\phantom{*}}}
{V_{td}^*V_{tb}^{\phantom{*}}},&
\displaystyle
\omega_{s_R} \equiv \frac{W_{\tilde{b}_Rs_R}^* W_{\tilde{b}_R b_R}^{\phantom{*}}}
{V_{td}^*V_{tb}^{\phantom{*}}}
\end{array}
\end{equation}
We expect that the various $\omega$ be complex numbers with
modulus and phase of order unity.

\item[ii)] The $\tilde{d}$ and $\tilde{s}$ squarks, $L$eft and $R$ight,
are degenerate to a high degree, with squared mass
$m^2_{12_L}$ and $m^2_{12_R}$,
whereas $\tilde{b}_L$ and $\tilde{b}_R$ have squared masses
$\mbL^2$ and $\mbR^2$ which can differ from $m_{12_{L,R}}^2$ by relative
order of unity.
Furthermore, it is more likely that $m_{\tilde{b}_{L,R}} < m_{12_{L,R}}$~\cite{3,6}.

\item[iii)] In the basis where the $\omega$ are defined,
there is still a small admixture between
the $\tilde{\rm d}_L$ and $\tilde{\rm d}_R$ squarks.
We assume, for simplicity, that the $A$-terms do not contain new flavour violations.
More precisely, for the scalar mixing terms we take the form
$$m_b(A_b+\mu\tan\beta)(\tilde{\rm d}_L^\dagger
W_{\tilde{\rm d}_L b_L}W_{ b_R\tilde{\rm d}_R}\tilde{\rm d}_R+
\hbox{h.c.}).$$
Being proportional to the relatively small $b$-quark mass,
we can treat these terms as perturbations.
\end{itemize}
Given this framework, loops of supersymmetric particles contribute with extra terms to
the CP-violating parameter in $K$-physics, $\varepsilon_K$, to the neutron EDM, $d_N$,
and to the $B$-$\bar{B}$ matrix elements of the effective Hamiltonian,
$M_{12}(B)\equiv\bAk{B^0}{{\cal H}_{\rm eff}}{\bar{B}^0}$.
Useful approximate formul\ae{} for these contributions are
\begin{eqnsystem}{sys:effects}
|\varepsilon_K |_{\rm SUSY}&\approx &
\frac{\alpha_3^2}{9\sqrt{2}}\frac{f_K^2 m_K^3}{m_s^2 \Delta m_K}
 |\omega_{d_L}\omega_{d_R}\omega_{s_L}^*\omega_{s_R}^*
(V_{td} V_{ts})^2| \frac{\eta_L\eta_R}{\max(\mbL^2,\mbR^2,M_3^2)}\\
d_N |_{\rm SUSY} &\approx & e\frac{2\alpha_3}{81 \pi} m_b
\hbox{Im}\,(\omega_{d_L}\omega_{d_R}) |V_{td}^2|
\frac{(A_b+\mu\tan\beta) M_3}{\max(\mbL^2,\mbR^2,M_3^2)^2}\eta_L\eta_R\\
M_{12}(B_d) |_{\rm SUSY}  &\approx & \frac{2\alpha_3^2}{9} f_{B_d}^2 m_{B_d}
(V_{td}^*V_{tb})^2 \left[
\frac{\omega_{d_L}^2\eta_L^2}{\max(\mbL^2,M_3^2)} +
\frac{\omega_{d_R}^2 \eta_R^2}{\max(\mbR^2,M_3^2)} +
4\frac{\omega_{d_L}\omega_{d_R} \eta_L\eta_R}{\max(\mbL^2,\mbR^2,M_3^2)} \right]
\label{eq:dmSUSY}\\
M_{12}(B_s) |_{\rm SUSY}  &\approx &M_{12}(B_d) |_{\rm SUSY}\hbox{ with $d\to s$}
\end{eqnsystem}
where we have used standard notations for the various quantities,
$M_3$ is the gluino mass, and
\begin{equation}
\eta_{L,R}\equiv 1 - \frac{m_{\tilde{b}_{L,R}}^2}{m_{12_{L,R}}^2}
\end{equation}
is a super-GIM suppression factor. At the unification scale we expect
$\eta^{\rm G}_{L,R}$ of order unity.
A large gluino mass, however, can significantly reduce
$\eta_{L,R}$ at the Fermi scale, since\footnote{In the formula for $\eta_L$
we have neglected, for simplicity, the $\lambda_t$-induced RGE effects.
They are in fact not very important since
the reduction of the $\tilde{b}_L$ mass is compensated by a corresponding
reduction of the ${\rm d}_L\tilde{\rm d}_L$ mixing angles.}
\begin{equation}
\eta_{L,R} \approx \frac{\eta_{L,R}^{\rm G}}{1 + 5.3 
(M_3^2/m_{12_{L,R}}^2)|_{\MGUT}}
\end{equation}
(``{\em gluino focusing\/}''),
a fact to be taken into account in realistic estimates of various effects~\cite{3}.

\begin{figure}[t]\setlength{\unitlength}{1cm}
\begin{center}
\begin{picture}(16,8.5)\Red
\put(3,8){fig.~\ref{fig:1}a: lighter gluino}
\put(12,8){fig.~\ref{fig:1}b: heavier gluino}\Black
\ifMac
{\put(-0.5,0){\special{picture SP300}}
\put(8.5,0){\special{picture SP500}}}
{\put(-0.5,0){\includegraphics{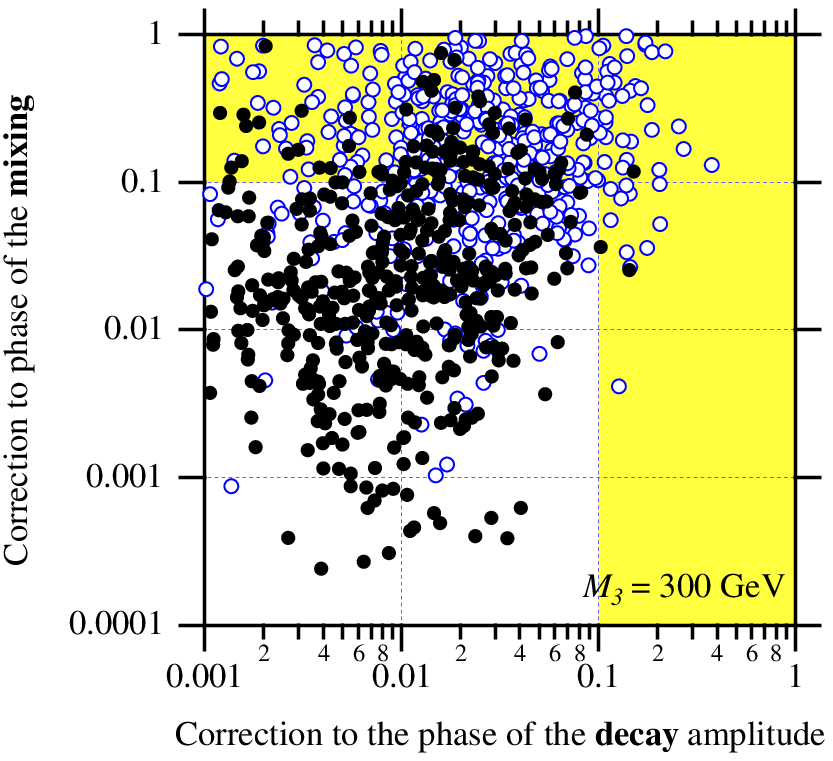}}
\put(8.5,0){\includegraphics{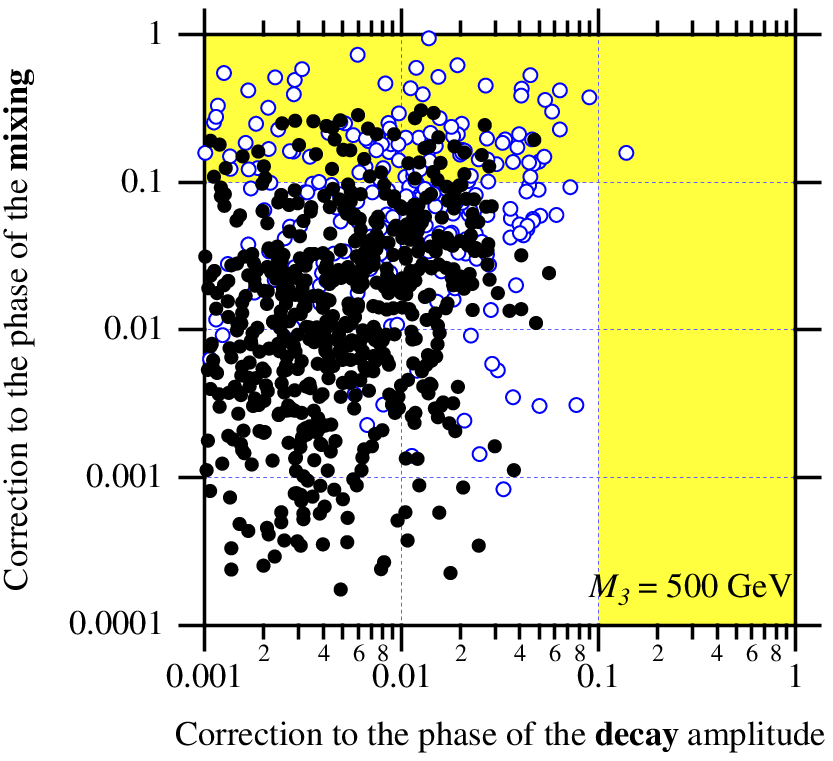}}}
%\put(-6.3,-2.4){\special{psfile=SPAM.ps}}  
%\put(7.4,0){\special{psfile=CPR.ps}}
\end{picture}
\caption[SP]{\em Scatter plots of the
supersymmetric corrections to the phases of the $B_d$ mixing, $2\varphi_{B_d}$, and of the
$B_d\to \phi K_S$ decay amplitude.
The empty points {\rm (`$\Blue\circ\Black$')} are excluded by too large supersymmetric
contributions to $\varepsilon_K$ or $d_N$.
The darker area is our estimate of where these effects could be detected in view of the
theoretical uncertainties.
\label{fig:1}}
\end{center}\end{figure}

\section{Supersymmetric effects in the decay amplitudes}
To study CP-violation in $B$-decays, one has to investigate the possible contributions
from supersymmetric loops not only to the $\Delta B=2$ mixing amplitudes but also
to direct decay amplitudes.
In general, this is not an easy task.
There are many possible different channels.
The SM amplitudes themselves are largely uncertain.
Finally, unlike the neutron EDM and the $\Delta S=2$ and
$\Delta B=2$ mixings, the $\Delta B=1$ non-leptonic amplitudes receive {\em a priori}
comparable contributions from three different kinds of supersymmetric loops:
``electric'' penguins, ``magnetic'' penguins and box diagrams.
Furthermore, to assess the possible size of these contributions, it is
essential to take into account the constraints coming from the other observables
mentioned above, which have either already
been measured ($\varepsilon_K$, $\Delta m_{B_d}$)
or are subject to stringent bounds (EDMs, $\mu\to e \gamma$, $\mu\to e$ conversion).

The interesting things to compare are supersymmetric loops with SM penguins~\cite{7}.
Anticipating the following discussion, we expect the second to dominate over the first.
As such, channels with negligible SM penguins, like $B_d\to\psi K_S$, will not be
affected by supersymmetric loops as well. On the contrary, one has to concentrate on
modes arising, in the SM, from pure loops, like those induced by $b\to s\bar{d}d$ and
$b\to s\bar{s}s$. Furthermore, $b\to s\bar{s}s$ is particularly interesting since in
the SM the corresponding amplitude only depends on one overall weak phase to a very
good approximation\footnote{A different possibility is given by $b\to s\gamma$
decays~\cite{Absg}. In the present framework, however, the bound on the neutron EDM
prevents a supersymmetric CP asymmetry in $b\to s\gamma$
significantly larger than the effect estimated in the SM.}.
Confining to the appendices a detailed description of the calculation
of the $b\to s\bar{s}s$ effective Hamiltonian and of the (uncertain) estimates
of the matrix elements for the $B_d\to\phi K_S$ decay amplitude,
we discuss the naive expectation for this amplitude.

In the SM, the amplitude for a $b\to s\bar{s} s$ decay has a parametric dependence on
the gauge couplings and on particle masses of the form
\begin{equation}\label{eq:7}
A^{\rm SM}\sim \frac{\alpha_2\alpha_3}{M_W^2}\ln\frac{m_t^2}{m_c^2}.
\end{equation}
The appearance of the charm mass in the argument of the logarithm in\eq{7}
reflects an infrared divergence in the SM penguin,
making its absolute estimate particularly uncertain.
At the same time, the supersymmetric contribution has the form
\begin{equation}\label{eq:8}
A^{\rm SUSY}\sim\frac{\alpha_3^2}{\max(m_{\tilde{b}}^2,M_3^2)}\eta
\end{equation}
where we do not distinguish between $L$ and $R$ in $m_{\tilde{b}}$ and $\eta$.
Eq.\eq{8} is appropriate for ``electric'' penguins,
but box diagrams and ``magnetic'' penguins
may give comparable contributions.

In an analogous way, the SM box diagrams for the $\Delta B=2$ mixings has the
parametric dependence
\begin{equation}\label{eq:9}
M_{12}^{\rm SM}\sim \frac{\alpha_2^2}{M_W^2}
\end{equation}
whereas, from eq.\eq{dmSUSY},
\begin{equation}\label{eq:10}
M_{12}^{\rm SUSY}\sim \frac{\alpha_3^2}{\max(m_{\tilde{b}}^2,M_3^2)}\eta^2
\end{equation}
Taking all $\omega$'s equal to unity,
eq.s~(\ref{eq:7}$\div$\ref{eq:10}) lead to the naive estimate
of the relative importance of supersymmetric corrections in decay amplitudes compared
to mixing as
\begin{equation}\label{eq:11}
\frac{A^{\rm SUSY}/A^{\rm SM}}{M_{12}^{\rm SUSY}/M_{12}^{\rm SM}}\sim
\frac{\alpha_2}{\alpha_3}\frac{1}{\eta\ln m_t^2/m_c^2}\sim
\frac{3\%}{\eta}
\end{equation}
Sizable supersymmetric contributions in the decay amplitudes appear less likely than
in the mixing.
Although the double ratio in\eq{11} increases when $\eta$ decreases, both
$A^{\rm SUSY}$ and $M_{12}^{\rm SUSY}$ gets suppressed in this case.

This naive expectation is confirmed by the
detailed calculation described in the appendices,
whose results are shown in figure~\ref{fig:1} in the form of scatter plots.
We compare the expected correction to the phase of the $B_d\to\phi K_S$ decay amplitude
with the correction to the phase of the $B_d$-$\bar{B}_d$ mixing.
In each graph the physical gluino mass is taken fix as indicated, as are the factors
$\eta$ at the unification scale,
$\eta^{\rm G}_L = \eta^{\rm G}_R = 0.75$
($m_{\tilde{b}}^{\rm G}= m_{12}^{\rm G}/2$).
On the contrary we allow variations of
the sbottom masses $\tilde{m}_b^{\rm G}$ in the range
$\mbR^{\rm G}=\mbL^{\rm G} = (\frac{1}{3}\div 3) M_3^{\rm G}$.
The Higgs soft masses are varied in the same range,
so that, taking a moderate $\tan\beta = (1.5\div 5)$ at the electroweak scale,
the $\mu$ term is determined
(up to its sign) from the minimization equations.
The $A$-term $A_b^{\rm G}$ is varied in the range $(-3\div 3) m_{{\tilde b}}^{\rm G}$.
Finally, all the $\omega$ factors at the unification scale
are varied between $\frac{1}{3}$ and $3$
in modulus with random phases\footnote{The resulting
effects are not different in minimal
SO(10) unification, even if this model does not give new
CKM-like phases in $\omega_{d_L}$ and $\omega_{s_L}$~\cite{6}.}.
The empty dots correspond to points of parameter space that give
rise to too large corrections to $\varepsilon_K$ and/or to the neutron electric dipole.
Possibly stronger bounds are obtained from
effects in the leptonic sector (like $\mu\to e\gamma$)
but a reliable comparison is possible only under specific
assumptions on the sfermion spectrum and the mixing angles.
Most of the points with small relative corrections both to mixing and decay
have a strong super-GIM suppression caused by ``gluino focusing''
(small $\eta$ at low energy).
The only possible enhancement of $A_{\rm SUSY}$ with respect to the naive estimate
might be due to ``magnetic'' penguins that give a
contribution proportional to $(A_b+\mu\tan\beta)/M_3$.
This factor cannot be too large, however, because it also enhances the neutron EDM.
As discussed below, the darker areas in fig.~\ref{fig:1} correspond to our estimate
of where these effects could be detected in view of the
theoretical uncertainties.

We have done an analogous study for the $B_s$-system,
which shows similar effects to the $B_d$ case, although
in general not strictly correlated to it.

\section{Measurements}
Suppose that supersymmetric loops indeed contribute to CP-violating
observables in $B$-decays. How can they be disentangled from pure
SM effects and how the relevant parameters can be finally measured?
The answers to these questions will crucially depend on the precise evolution of the
experimental programme, which is difficult to anticipate.
Furthermore, these same questions have
already attracted interest in the literature~\cite{7,8}.
Still, we find useful to summarize our view on this issue, having in mind the special
physical situation outlined above.

As customary~\cite{9,10}, we define two new phases, $\varphi_{B_d}$ and $\varphi_{B_s}$ as
\begin{equation}\label{eq:12}
M_{12}^{\rm SM}(B_{d,s}) + M_{12}^{\rm SUSY}(B_{d,s}) =
M_{12}^{\rm SM}(B_{d,s}) r_{d,s} \exp[2i\varphi_{B_{d,s}}]
\end{equation}
with $r_{d,s}$ real positive numbers.
These phases have direct physical meaning, as the usual angles
$\alpha$, $\beta$, $\gamma$ of the CKM unitarity triangle ($\alpha+\beta+\gamma=\pi$),
being independent from any phase convention for the various fields.

A first point, already mentioned,
is that the $B$-decay amplitudes largely dominated in
the SM by tree level transitions will not be affected by supersymmetric corrections.
In particular the CP-asymmetries
\begin{equation}
\frac{\Gamma(B^0(t)\to f)-\Gamma(\bar{B}^0(t)\to f)}
{\Gamma(B^0(t)\to f)+\Gamma(\bar{B}^0(t)\to f)} = -a(B\to f)\sin\Delta m_B t,
\qquad\Delta m_B>0,
\end{equation}
for the modes $B_d\to\psi K_S$ or $B_s\to \psi\phi$
\footnote{The $B_s$ system is different from $B_d$ because of the large mixing and the
non negligible $\Delta\Gamma/\Gamma$, which is ignored here.
In view of this, untagged rates might be
more useful to extract $\sin2\varphi_{B_s}$ from
$B_s\to\psi\phi$~\cite{dGamma}.} will measure well defined combinations
of CP-phases
\begin{eqnsystem}{sys:14}\label{eq:14a}
&&a(B_d\to \psi K_S) =\sin 2(\beta+\varphi_{B_d}),\\
&&a(B_s\to \psi \phi|_{{\rm CP}=-}) =\sin 2\varphi_{B_s}.
\label{eq:ABs}
\end{eqnsystem}
While\eq{ABs} could allow a direct measurement of $\varphi_{B_s}$,
the study of such `clean' $B$-decays alone,
however, will not allow the extraction of the
CKM angles $\alpha$, $\beta$, $\gamma$,
unless supplemented with independent
informations on the lengths of the sides of the unitarity triangle,
i.e.\ on $|V_{ub}|$ and $|V_{td}|$~\cite{9}.
Some $B$-decays may provide this information, but only if hadronic uncertainties
are kept under control\footnote{Independent clean information~\cite{Burassone} could
come from $K^+\to \pi^+ \nu\bar{\nu}$ and
$K_L\to \pi^0\nu\bar{\nu}$ which are not affected by the supersymmetric
corrections considered in this paper~\cite{5}.}.
Notice on the contrary that the ratio of the mixing parameters
\begin{equation}\label{eq:15}
\left|\frac{M_{12}(B_d)}{M_{12}(B_s)}\right| = \frac{r_d}{r_s}
\left|\frac{V_{td}}{V_{ts}}\right|^2,
\end{equation}
up to perhaps controllable SU(3)-breaking effects, will involve an extra unknown,
$r_d/r_s$.

\medskip

To go further, one would like to check if the
supersymmetric corrections to the decay amplitudes
are indeed relatively suppressed,
as indicated by the calculation described above.
A sign that they are not, would be a detection of a large
difference in appropriate
asymmetries, sufficiently free from theoretical uncertainties, like
%\begin{eqnsystem}{sys:16}\label{eq:phiKS}
\begin{equation}\label{eq:16}
a(B_d\to \phi K_S) \neq a(B_d\to \psi K_S).
\end{equation}
%a(B_d\to \pi^0 K_S) &\neq& a(B_d\to \psi K_S).\\[2mm]
%\riga{
On the basis of the previous calculation, we expect a phase difference
$\delta\varphi$ between the phases entering the
left and right hand sides of~(\ref{eq:16}) %(\ref{sys:16}a,b)
not exceeding 0.2.
The major uncertainty in determining the maximal effect in $\delta\varphi$
is represented by the estimate of the matrix element of the magnetic penguin operator.
In the case of\eq{16}, %\eq{phiKS},
however, a phase difference $\delta\varphi\circa{>}0.1$ would be
sufficient to indicate an effect beyond the SM.\\
%Alternatively, one could look for a difference in charged
%$B$-decay widths into CP conjugate modes}\\
%\Gamma(B^+\to \pi^+ K^0) &\neq&\Gamma(B^- \to \pi^- \bar{K}^0).
%\end{eqnsystem}
%According to our estimate,
%a maximal difference in the strong phases between the SM and the supersymmetric
%contribution could give rise to a up to $30\%$ difference between these two rates,
%while a difference larger than $1\%$ is
%sufficient to indicate an effect beyond the SM~\cite{B+B-}.

\smallskip

Let us suppose that we can neglect supersymmetric corrections in decay amplitudes.
How could one then proceed in order to measure the angles $\alpha$, $\beta$, $\gamma$
and $\varphi_{B_d}$, $\varphi _{B_s}$?
Let us also take the point of view that the various hadronic uncertainties will
deteriorate the experimental informations obtainable on the sides of the unitarity
triangle in a significant way. Although this may be pessimistic, the problem of the
theoretical uncertainties is an obvious matter of concern if one wants to extract new
effects from the experimental data.

Although not at the first step of a $B$-factory, the measurements of
$B_d\to \pi\pi$, with an appropriate isospin analysis~\cite{12}, and of
$B\to DK$~\cite{13} will give access to $\alpha-\varphi_{B_d}$ and $\gamma$
respectively. Notice however that {\em even these informations, together with\eq{14a},
are not enough to separate $\alpha$, $\beta$ and $\varphi_{B_d}$}.

The possibility that has been considered in the context of the SM is a cumulative
study~\cite{14,15,16} of
$B\to\pi\pi$, $\pi K$ and $KK$.
In the present situation we would summarize this possibility as follows.
Of special interest are the decays $B_d^0\to \pi^+\pi^-$,
$B^+\to \pi^+ K^0$, $B^0\to \pi^- K^+$ and their charge conjugate.
It is useful to parametrize the corresponding amplitudes as
\begin{eqnsystem}{sys:19}
A(B_d^0\to \pi^+\pi^-) &=& T^\pi + P_d\\
A(B^+\to \pi^+K^0) &=& P_s\\
A(B_d^0\to \pi^-K^+) &=& T^K + P_s
\end{eqnsystem}
where, to a good accuracy, $T$ represents tree level contributions whereas $P$
arises from gluonic penguins\footnote{``Annihilation''
diagrams and electro-weak penguins are
not included in\sys{19}. They are unlikely, however,
to play any significant role~\cite{15,16,17}.}.
Defining
\begin{eqnsystem}{sys:20}
\noalign{\hbox{\parbox{\textwidth}{
~\hfill $T^\pi = V_{ud} V_{ub}^* t^\pi,\qquad T^K = V_{us} V_{ub}^* t^K,$\hfill~}}}\\
P_d &=& V_{td} V_{tb}^* p_d + V_{cd} V_{cb}^* \Delta p_d,\\
P_s &=& V_{ts} V_{tb}^* p_s + V_{cs} V_{cb}^* \Delta p_s,
\end{eqnsystem}
one gets the amplitudes for the charge conjugate modes by complex conjugation of the
CKM matrix elements $V_{ij}$, whereas the complex amplitudes $t$, $p$, $\Delta p$
remain unchanged.

The measurements of $\Gamma(B^\pm\to \pi^\pm K^0)$,
$\Gamma(B_d^0\to \pi^- K^+)$,
$\Gamma(\bar{B}_d^0\to \pi^+ K^-)$,
$\Gamma(B_d^0(t)\to \pi^+\pi^-)$ and $\Gamma(\bar{B}_d^0(t)\to \pi^+\pi^-)$
give six relations (five widths and $a(B_d\to \pi^+\pi^-)$) among the
amplitudes $T$, $P$ and the mixing phase $\beta+\varphi_{B_d}$.
In general, however, even taking into account that $t^\pi$ and $t^K$ can be taken
real by phase redefinitions, this is not enough to
constrain the many more unknowns in\sys{20}.
To get such constraints it is possible to explore 
the SU(3) relations among the different
amplitudes and/or, but this may be more risky~\cite{15}, the fact that
$|\Delta p/p|$ could be sufficiently small.

In the exact SU(3) limit one has
\begin{eqnsystem}{sys:}
t^\pi = t^K &\equiv& |t|\\
p_d = p_s &\equiv& |p| e^{i\delta_p}\\
\Delta p_d = \Delta p_s &\equiv& |\Delta p| e^{i\delta_{\Delta p}},
\end{eqnsystem}
i.e.\ five real parameters.
To the extent that SU(3) violations maybe accounted for~\cite{16}
without introducing new unknowns, the programme of measuring all the CP phases
could be successfully accomplished.
Measurements of the suppressed modes $B\to KK$ will also provide
additional information~\cite{15}.

Altogether, it seems possible to extract $\varphi_{B_d}$ from the data
in different ways, all involving some amount of hadronic uncertainty.

\section{Conclusions}
In conclusion, we have studied the possible deviations
from the expectation of the standard model
in signals of CP-violation in hadronic $B$ decays,
as they occur in a definite and well
motivated framework for flavour physics and supersymmetry.
Such deviations can arise either though extra phases in $B_d$-$\bar{B}_d$
and $B_s$-$\bar{B}_s$ mixing, or in the decay amplitudes.
To assess both the absolute and the relative importance of these new effects,
we have considered
the constraints coming from other flavour observables either already measured
or subject to strong bounds.

We find that visible effects are possible in a portion of the parameter space.
To make them manifest, however, it will be essential to study appropriate
observables with minimal contamination from various theoretical
uncertainties, as already suggested by several authors.

\appendix\small

\setcounter{equation}{0}
\renewcommand{\theequation}{\thesection.\arabic{equation}}

\section{Supersymmetric corrections to mixing-induced CP violation}
The SM contribution to the $B^0_q$-$\bar{B}^0_q$ mixing ($q=\{d,s\}$) is
\begin{eqnsystem}{sys:dmB}
% attenzione che M_12 = Delta m / 2  per cui  (fB^2 mB/3) <---> (2/3  fB^2 mB)
M_{12}(B_q)|_{\rm SM} &=& \frac{\alpha_2^2}
{8 M_W^2}(\frac{1}{3} f_B^2 m_B) (V_{tq}^*V_{tb})^2
\eta^4 2.5\\
\riga{Here $\eta = (\alpha_3(M_Z)/\alpha_3(m_b))^{3/46}\approx 0.954$ gives the leading order QCD
renormalization effect
and we have computed the matrix elements in the vacuum saturation approximation.
In the same approximation the supersymmetric gluino correction is given by}\\
\nonumber
M_{12}(B_q)|_{\rm SUSY} &=& \frac{\alpha_3^2}{9M_3^2}
(\frac{1}{3} f_B^2 m_B) (V_{tq}^*V_{tb})^2\bigg[
\eta^4\omega_{q_L}^2 \{ 11 \Bk + \BM\}(\{\rbL-\rdL\},\{\rbL-\rdL\})+\\
&&+\omega_{q_L}\omega_{q_R}
\{\frac{20\eta^2-14\eta^{-16}}{3}\Bk+\frac{\eta^2+56\eta^{-16}}{3}\BM\}(\{\rbL-\rdL\},\{\rbR-\rdR\}) +\\
&&+\eta^4\omega_{q_R}^2 \{ 11 \Bk + \BM\}(\{\rbR-\rdR\},\{\rbR-\rdR\})\bigg]
\nonumber\end{eqnsystem}
where, in order to avoid long expressions, we have
introduced the following compact notations:
$$\rbL \equiv \frac{\mbL^2}{M_3^2},\qquad
\rdL \equiv \frac{m_{12_L}^2}{M_3^2},\qquad
\rbR \equiv \frac{\mbR^2}{M_3^2},\qquad
\rdR \equiv \frac{m_{12_R}^2}{M_3^2}$$
and
$$f(\{a_1\pm a_2\})\equiv f(a_1) \pm  f(a_2),\qquad
\{f_1\pm f_2\}(a) \equiv f_1(a) \pm f_2(a)$$
The various loop functions are defined below in eq.s\sys{Box}.
The QCD corrections have been taken from~\cite{SUSY-QCD-RGE}.

The other supersymmetric contribution are not considered here.
The ones mediated by charged Higgs and by the higgsino component of the charginos
have the same phase and angles of the SM contribution and smaller size.
The ones mediated by the weak gauginos are proportional to the new mixing parameters
$\omega$ but have smaller vertex factors.

\setcounter{equation}{0}

\section{Supersymmetric corrections to direct CP violation}\label{Adecay}
The dominant {\bf SM contribution} to the effective Hamiltonian
for $b\to s\bar{s} s$, recalled here to estabilish
the notation and to define our approximations,
is well known~\cite{Burassone} and
is given by ``electric'' penguins, that
are enhanced by an infrared logarithmic factor, $\ln M_W^2/m_c^2$.
For this reason there are two significant
contributions from top and charm loops.
Since $V_{cb}V_{cs}^*\approx - V_{tb}V_{ts}^*$
these two contributions have the same CKM phase
so that the phase of the SM decay amplitude is well predicted.
We also include ``magnetic penguins'', not
enhanced by infrared divergences in the limit $m_c\to 0$.
We neglect electroweak penguins and box
diagrams that are suppressed by powers of $\alpha_2/\alpha_3$.

The inclusion of QCD effects
from the $M_W$ scale to the $m_b$ scale
is done in the usual effective field theory approach.
At the $M_W$ scale the heavy particles
%($W$ boson, top quark, SUSY particles)
are integrated out
and the eight current-current operators,
\begin{eqnsystem}{sys:Ops}
\Opi{AB} &\equiv&  (\bar{s}^i\gamma_\mu {\cal P}_A b^i)
\sum_q(\bar{q}^j\gamma_\mu {\cal P}_B q^j), \\
\Opx{AB} &\equiv& (\bar{s}^i\gamma_\mu {\cal P}_A b^j)
\sum_q(\bar{q}^j\gamma_\mu {\cal P}_B q^i),\\
\riga{the two cromo-magnetic penguin operators,}\\[-3mm]
\label{eq:O'}
{\cal O}'_A&\equiv&  m_bg_3(\bar{s}\gamma_{\mu\nu}T^a {\cal P}_A b)G_{\mu\nu}^a,\\
\riga{and the `charm operators',}\\[-3mm]
{\cal O}_c^= &=& (\bar{s}^i\gamma_\mu {\cal P}_L c^i)
(\bar{c}^j\gamma_\mu {\cal P}_L b^j)\\
{\cal O}_c^\times &=& (\bar{s}^i\gamma_\mu {\cal P}_L c^j)
(\bar{c}^j\gamma_\mu {\cal P}_L b^i),
\end{eqnsystem}
have to be included in the effective Hamiltonian
\begin{equation}
{\cal H}_{\rm eff}^{\em SM}(b\to s\bar{s}s) = -\frac{g_2^2}{2M_W^2}V_{tb} V_{ts}^*
(C_c^= {\cal O}_c^= + C_c^\times {\cal O}_c^\times+
C_{LL}^= \Opi{LL} + C_{LL}^\times \Opx{LL} +
C_{LR}^= \Opi{LR} + C_{LR}^\times \Opx{LR}+
C'_{R} {\cal O}'_{R}).
\end{equation}
In these equations $i,j$ are color indices and $A,B = \{L,R\}$.
At the electroweak scale $C_c^= =1$, $C_c^\times=0$ and $C_{LA}=\Ord(\alpha_3/4\pi)$
while operators with a right-handed $s$-quark have negligible coefficients.
At the $m_b$ scale the Wilson coefficients $C_{LA}$
of the current-current operators ${\cal O}_{LA}$ get an infrared enhancement
due to the mixing with the charm operators.
These coefficients are thus of order
$\ell\equiv (\alpha_3/4\pi) \ln M_W^2/m_b^2 \approx 0.1$
or, more precisely,
\begin{equation}
\begin{array}{lll}
C_c^==+1.1,\qquad &C_{LL}^=   = +0.014,\qquad &C_{LR}^=   = +0.009,\\
C_c^\times=-0.3,&C_{LL}^\times = -0.031,&C_{LR}^\times  =-0.040,
\end{array}\qquad
C'_{R}=-\frac{0.15}{(4\pi)^2}.
\end{equation}
In the RGE evolution we have kept the leading-order terms of order
$\ell^n$ and the terms of order $(\alpha_3/4\pi)$.
The smaller next-to-leading-order terms of order
$(\alpha_3/4\pi)\ell^n$ (with $n\ge 1$)
(that we have systematically neglected)
contain all the technical problems typical of a full NLO computation.
At the $m_b$ scale we can finally compute the charm loop
in the effective theory, getting
a further contribution to the $C_{LA}$ coefficients
of order $\alpha_3(m_b)/(4\pi)$.
Computing this charm loop at the hadronic level
could give a larger, and maybe dominant,
contribution to the decay amplitude~\cite{martinelli}.
Here we compute the charm loop
at quark level, so that we have ${\rm B.R.}(B_d\to K_S\phi)\approx 0.8\cdot 10^{-5}$.

\medskip

We now evaluate the {\bf supersymmetric contributions}. In the supersymmetric case,
``electric'' gluino penguins (`pengluins') are not enhanced by an infrared logarithm so
that electric pengluins do not dominate over box diagrams and magnetic pengluins, that
have comparable vertex and loop factors. Since at hadronic level the relevant decays
involve scalar or vector particles, all the various amplitudes with different quark
helicity structure interfere with the SM one.

In order to include the perturbative QCD corrections at scales between $m_t$ and $m_b$,
we write the effective Hamiltonian for the $b\to s\bar{s}s$ decay as
$${\cal H}_{\rm eff}^{\rm SUSY}(b\to s\bar{s}s) = -V_{ts}^*V_{tb}
\bigg[\sum_{AB} \frac{\alpha_3^2}{M_3^2}(c_{AB}^=\Opi{AB}+c_{AB}^\times\Opx{AB})+
\frac{\alpha_3}{4\pi M_3^2} (c'_{R} {\cal O}'_{R}+c'_{L} {\cal O}'_{L})\bigg]$$
where the sum now extends over all the
eight current-current operators, defined in eq.s~(\ref{sys:Ops}).
The gluino-mediated supersymmetric contributions
to the Wilson coefficients $c$ at the weak scale are
%proportional to the new CP violating mixings $\omega$ are
\begin{eqnsystem}{sys:coefficienti}
c_{LL}^=  &=&   +\frac{1}{2}\omega_{s_L}
\Pglu(\{\rbL - \rdL\})+\omega_{s_L}
\frac{1}{9}\{5\BM+21 \Bk\}(\{\rbL-\rdL\},\rdL),\\
c_{LL}^\times  &=&  -\frac{3}{2}\omega_{s_L}
\Pglu(\{\rbL - \rdL\})+\omega_{s_L}
\frac{1}{9}\{-3\BM+\Bk\}(\{\rbL-\rdL\},\rdL),\\
c_{LR}^=  &=&  +\frac{1}{2}\omega_{s_L}
\Pglu(\{\rbL - \rdL\})+\omega_{s_L}
\frac{1}{18}\{20\BM+21\Bk\}(\{\rbL-\rdL\},\rdR),\\
c_{LR}^\times  &=&  -\frac{3}{2}\omega_{s_L}
\Pglu(\{\rbL - \rdL\})+\omega_{s_L}
\frac{1}{18}\{\BM-12\Bk\}(\{\rbL-\rdL\},\rdR).\\
\riga{The coefficients for the current-current operators ${\cal O}_{{RA}}$
with a right-handed bottom quark
are obtained interchanging $L\rightleftharpoons R$,
$\omega_{s_L}\rightleftharpoons\omega_{s_R}$ and $\ell\rightleftharpoons r$
in the previous expressions.
The coefficients of the magnetic penguin operators are}\\
c'_{R} &=& \frac{3}{2}\omega_{s_L} \left[|V_{\tilde{b}_R b_R}|^2
\frac{A_b+\mu\tan\beta}{M_3}
\PgluI(\{\rbL-\rdL\},\rbR) + \PgluE(\{\rbL-\rdL\})\right],\\
c'_{L} &=& \frac{3}{2}\omega_{s_R} \left[ |V_{\tilde{b}_L b_L}|^2
\frac{A_b+\mu\tan\beta}{M_3}
\PgluI(\{\rbR-\rdR\},\rbL)\right].
\end{eqnsystem}
For simplicity, we have given their expressions in the limit where the $A$-terms
are generation universal at the Fermi scale.
The expressions are easily modified if $A_d=A_s\neq A_b$.

The inclusion of QCD effects is well known~\cite{Burassone}.
The coefficients of the two cromo-magnetic
penguins renormalize multiplicatively as
$c'_A(m_B) = \eta^{-14/23}  c'_A(M_Z)$, where $A = \{L,R\}$
and $\eta$ has been defined as $\alpha_3(M_Z)/\alpha_3(m_b)$.
The four current-current operators with a left-handed $b$-quark mix among themselves.
The other four current-current operators with a
right-handed $b$-quark also mix among themselves,
with the mixing described by the same mixing matrix.

\medskip

We finally list all the loop functions used above.
The penguin loop functions that appear in the supersymmetric contributions are
\begin{equation}\label{eq:Penguins}
\Pglu = P_F - \frac{1}{9} P_B,\qquad
\PgluE = P_{FE} - \frac{1}{9} P_{BE},\qquad
\PgluI = P_{FI} - \frac{1}{9} P_{BI}
\end{equation}
where
\begin{eqnsystem}{sys:Box}
P_F(r) &=& \frac{1}{36}\frac{1}{(r-1)^4}[7-36r+45r^2-16r^3-(18-12r)r^2\ln r]\\
P_B(r) &=& \frac{-1}{36}\frac{1}{(r-1)^4}[11-18r+9r^2-2r^3+6\ln r]\\
P_{FE}(r) &=&  \frac{1}{12}\frac{-1}{(r-1)^4}[1-6r+3r^2+2r^3-6r^2\ln r]\\
P_{BE}(r) &=&  \frac{1}{12}\frac{1}{(r-1)^4}[2+3r-6r^2+r^3+6r\ln r]\\
P_{FI}(r) &=&  \frac{1}{2}\frac{-1}{(r-1)^3}[4r-1-3r^2+2r^2\ln r] \\
P_{BI}(r) &=& \frac{1}{2}\frac{1}{(r-1)^3}[r^2-1-2r\ln r] \\
\riga{The box loop functions are defined as}\\[-3mm]
\Bk(r_1,r_2) &=&  i(4\pi)^2\int \frac{d^4 k}{(4\pi)^4}
\frac{{1\over 4} M^2 k^2}{(k^2-M^2)^2(k^2-r_1^2M^2)(k^2-r_2^2M^2)}\\
\BM(r_1,r_2) &=&  i(4\pi)^2\int\frac{d^4 k}{(4\pi)^4}
\frac{M^4}{(k^2-M^2)^2(k^2-r_1^2M^2)(k^2-r_2^2M^2)}
\end{eqnsystem}
where $M$ is an arbitrary mass scale.

\setcounter{equation}{0}

\section{Hadronic matrix elements for $b\to s\bar{s}s$ decays}
We give here all the relevant hadronic matrix
elements computed in the vacuum-insertion approximation.

Neglecting the `annihilation diagrams',
that have a strongly suppressed form factor~\cite{charmless},
all the necessary matrix elements can be expressed in terms of
\begin{eqnarray*}
\bAk{\phi (p_\phi ,\epsilon_\phi )}{\bar{s}\gamma_\mu s}{0} &=&
f_\phi  m_\phi^2 \varepsilon_\mu^\phi,  \\
\bAk{K^0(p_K)}{\bar{s}\gamma_\mu b}{\bar{B}_d^0(p_B)} &=& 
F_+^{BK}(t) (p_B+p_K)_\mu + F_-^{BK}(t) (p_B-p_K)_\mu,
\end{eqnarray*}
where $t\equiv (p_B-p_K)^2$.
Matrix elements of scalar operators are obtained using the equations of motion for the
quark fields. Matrix elements of tensor operators, like $\bAk{K}{\bar{s}\gamma_{\mu\nu}
b}{B}$, are obtained using the heavy-quark effective theory~\cite{HQET} while
$\bAk{\phi}{\bar{s}\gamma_{\mu\nu}s}{0}$ can be estimated using a quark model as
in~\cite{magnetic}.
%The values of the form-factors are
%$f_\phi = -0.228$, $F_+^{BK}(0) =0.38 $
%and $F_+^{BK}(q^2) =F_+^{BK}(0)/(1-q^2/m_{B^*}^2) $ is dominated by a $B^*$ pole.
%The $F_-$ are obtained from the constituent U(2,2) quark model~\cite{U(22)}.
In terms of a single combination
of form-factors~\cite{charmless,BSW},
$H\equiv  (p_B\cdot\varepsilon_\phi) 2f_\phi  m_\phi^2  F_+^{BK}(m_\phi^2)$,
that cancels in the CP-asymmetries, we obtain
\begin{eqnsystem}{sys:B->Kphi}
\bAk{\phi K_S}{\Opi{LL}}{\bar{B}_d^0}  &=&\bAk{\phi K_S}{\Opi{RR}}{\bar{B}_d^0} =
\frac{1}{4}H(1+\frac{1}{N_c}) \\
\bAk{\phi K_S}{\Opx{LL}}{\bar{B}_d^0}  &=&
\bAk{\phi K_S}{\Opx{RR}}{\bar{B}_d^0} =\frac{1}{4}H(1+\frac{1}{N_c}) \\
\bAk{\phi K_S}{\Opi{LR}}{\bar{B}_d^0}  &=&\bAk{\phi K_S}{\Opi{RL}}{\bar{B}_d^0} =
\frac{1}{4} H\\
\bAk{\phi K_S}{\Opx{LR}}{\bar{B}_d^0}  &=&\bAk{\phi K_S}{\Opx{RL}}{\bar{B}_d^0}  =
\frac{1}{4}H\frac{1}{N_c} \\
\bAk{\phi K_S}{{\cal O}'_{L}}{\bar{B}_d^0}&=&\label{eq:<cromo>}
\bAk{\phi K_S}{{\cal O}'_{R}}{\bar{B}_d^0} \approx (1.2\pm 0.4) 
g_3^2 H\frac{N_c^2-1}{2N_c^2}
\end{eqnsystem}
where $N_c$ is the number of colors.
The computation of the matrix elements of the current-current
operators is straightforward.
More lengthy is the computation of the matrix element of the cromo-magnetic penguin
operator. We have estimated it in the following way.
To begin with, the cromo-magnetic penguin of eq.\eq{O'}
has been converted into a four-quark operator attaching a $\bar{s}s$ pair to the gluon.
Since the $\bar{s}s$ pair is in an octet state, in the factorization approximation
only the Fierz transformed operator contributes.
Simplifying the operator using the Dirac equation we get
\begin{eqnarray*}
\bAk{\phi K_S}{{\cal O}'_{L,R}}{\bar{B}_d^0}&=&
\frac{g_3^2}{k^2}\frac{N_c^2-1}{2N_c^2}\frac{1}{4}
\bigg\{ 2m_b^2\bAk{\phi}{\bar{s} \gamma_\mu  s}{0}
\bAk{K_S}{\bar{s}\gamma_\mu  b}{\bar{B}_d^0}+ \\
&&+m_b (p_b+p_{s_\phi})_\mu\times\big[
\bAk{\phi}{\bar{s}\gamma_\mu s}{0}\bAk{K_S}{\bar{s} b}{\bar{B}_d^0} +\\
&&
+i\bAk{\phi}{\bar{s}\gamma_{\mu\nu} s}{0}\bAk{K_S}{\bar{s}\gamma^\nu  b}{\bar{B}_d^0}
-i\bAk{\phi}{\bar{s}\gamma^\nu s}{0}\bAk{K_S}{\bar{s}\gamma_{\mu\nu} b}{\bar{B}_d^0}
\big]  \bigg\}.
\end{eqnarray*}
We notice that the first ordinary current-current term
gives the largest contribution to the full cromo-magnetic hadronic matrix element,
so that the numerical coefficient in eq.\eq{<cromo>}
is not very uncertain.
Adding the contributions from the other terms, and estimating the gluon momentum
$k^2\approx\frac{1}{2} (m_B^2 - \frac{1}{2} m_\phi^2 + m_K^2)$
assuming that each one of the two $s$ quarks inside the $\phi$
carry similar momentum $p_\phi/2$~\cite{magnetic},
we obtain the estimate of eq.\eq{<cromo>}.
%As a byproduct, we estimate that the SM-cromomagnetic penguin
%operator gives a $\sim30\%$ correction
%to the branching ratio $\hbox{B.R.}(B_d\to\phi  K_S)$,
%dominated by current-current operators.
Working in the factorization approximation,
we have neglected $\phi$ hadronization from $\bar{s}s$ quarks in octet color state,
that is kinematically enhanced by a smaller transferred momentum,
$k^2\approx m_\phi^2$.
This same kinematical factor enhances also the contribution from photonic magnetic
penguins, that remain however unimportant due to the small d-quark electric charge.

\Blue
\newpage

\small~

\end{document}
\\
Title: CP violation in B decays and supersymmetry
Authors: Riccardo Barbieri and Alessandro Strumia
Comments: 10 pages
Report-no: IFUP-TH 16/97
\\
CP violation in hadronic B-decays is studied in a definite and
well motivated framework of flavour physics and supersymmetry.
Possible deviations from the standard model both
in mixing and in decay amplitudes are discussed.
An attempt is made to describe an experimental strategy for looking
at these deviations and for measuring the relevant parameters.
\\